\documentclass[prl,superscriptaddress,showpacs,byrevtex,twocolumn]{revtex4}

\usepackage{graphicx}
\usepackage{epsfig}
\usepackage{dcolumn}
\usepackage{bm}
\usepackage{color}
\usepackage{ulem}

\newcommand{\gev}{\rm GeV}
\newcommand{\gevc}{{\rm GeV}/c}
\newcommand{\gevcs}{{\rm GeV}/c^2}
\newcommand{\mev}{\rm MeV}
\newcommand{\mevcs}{{\rm MeV}/c^2}
\newcommand{\kev}{\rm keV}

\newcommand{\ev}{\rm eV}

\newcommand{\MMS}{M_{\rm rec}^2}

\newcommand{\lum}{{\cal L}}
\newcommand{\eff}{\varepsilon}
\newcommand{\BR}{{\cal B}}

\newcommand{\pip}{\pi^+}
\newcommand{\pim}{\pi^-}
\newcommand{\piz}{\pi^0}

\newcommand{\psp}{\psi(2S)}

\newcommand{\jpsi}{J/\psi}
\newcommand{\psift}{\psi(4040)}
\newcommand{\psifto}{\psi(4160)}
\newcommand{\psiftf}{\psi(4415)}
\newcommand{\EE}{e^+e^-}
\newcommand{\MM}{\mu^+\mu^-}
\newcommand{\LL}{\ell^+\ell^-}
\newcommand{\GG}{\gamma\gamma}

\newcommand{\ppp}{\pi^+\pi^-\pi^0}

\newcommand{\etajpsi}{\eta J/\psi}
\newcommand{\ppjpsi}{\pi^+\pi^- J/\psi}

\newcommand{\pppsp}{\pi^+\pi^- \psp}

\newcommand{\reduline}{\bgroup\markoverwith
{\textcolor{red}{\rule[0.5ex]{2pt}{0.4pt}}}\ULon}

\newcommand{\beq}{\begin{equation}}
\newcommand{\eeq}{\end{equation}}
\newcommand{\bitm}{\begin{itemize}}
\newcommand{\eitm}{\end{itemize}}

\def\Journal#1#2#3#4{{#1} {\bf #2}, #3 (#4)}

\def\NIMA{Nucl. Instrum. Methods A}

\def\PRL{Phys. Rev. Lett.}
\def\PRD{Phys. Rev. D}

\def\EPJC{Eur. Phys. J. C}

\begin{document}

\preprint{} \preprint{ \vbox{ \hbox{   }
        \hbox{Intended for {\it Phys. Rev. Lett.}}
        \hbox{Authors: X. L. Wang, Y. L. Han, C. Z. Yuan, C. P. Shen, P. Wang}
        \hbox{Committee: P. Pakhlov (chair), S. Uehara, P. Lukin}
        }}
\title{
Observation of $\psift$ and $\psifto$ decay into $\eta\jpsi$}

\affiliation{Budker Institute of Nuclear Physics SB RAS and Novosibirsk State University, Novosibirsk 630090}
\affiliation{Faculty of Mathematics and Physics, Charles University, Prague}
\affiliation{University of Cincinnati, Cincinnati, Ohio 45221}
\affiliation{Justus-Liebig-Universit\"at Gie\ss{}en, Gie\ss{}en}
\affiliation{Gyeongsang National University, Chinju}
\affiliation{Hanyang University, Seoul}
\affiliation{University of Hawaii, Honolulu, Hawaii 96822}
\affiliation{High Energy Accelerator Research Organization (KEK), Tsukuba}
\affiliation{Indian Institute of Technology Guwahati, Guwahati}
\affiliation{Indian Institute of Technology Madras, Madras}
\affiliation{Institute of High Energy Physics, Chinese Academy of Sciences, Beijing}
\affiliation{Institute of High Energy Physics, Vienna}
\affiliation{Institute of High Energy Physics, Protvino}
\affiliation{INFN - Sezione di Torino, Torino}
\affiliation{Institute for Theoretical and Experimental Physics, Moscow}
\affiliation{J. Stefan Institute, Ljubljana}
\affiliation{Kanagawa University, Yokohama}
\affiliation{Institut f\"ur Experimentelle Kernphysik, Karlsruher Institut f\"ur Technologie, Karlsruhe}
\affiliation{Korea Institute of Science and Technology Information, Daejeon}
\affiliation{Korea University, Seoul}
\affiliation{Kyungpook National University, Taegu}
\affiliation{\'Ecole Polytechnique F\'ed\'erale de Lausanne (EPFL), Lausanne}
\affiliation{Faculty of Mathematics and Physics, University of Ljubljana, Ljubljana}
\affiliation{Luther College, Decorah, Iowa 52101}
\affiliation{University of Maribor, Maribor}
\affiliation{Max-Planck-Institut f\"ur Physik, M\"unchen}
\affiliation{University of Melbourne, School of Physics, Victoria 3010}
\affiliation{Graduate School of Science, Nagoya University, Nagoya}
\affiliation{Nara Women's University, Nara}
\affiliation{National United University, Miao Li}
\affiliation{Department of Physics, National Taiwan University, Taipei}
\affiliation{H. Niewodniczanski Institute of Nuclear Physics, Krakow}
\affiliation{Nippon Dental University, Niigata}
\affiliation{Niigata University, Niigata}
\affiliation{University of Nova Gorica, Nova Gorica}
\affiliation{Osaka City University, Osaka}
\affiliation{Pacific Northwest National Laboratory, Richland, Washington 99352}
\affiliation{Peking University, Beijing}
\affiliation{Research Center for Electron Photon Science, Tohoku University, Sendai}
\affiliation{University of Science and Technology of China, Hefei}
\affiliation{Seoul National University, Seoul}
\affiliation{Sungkyunkwan University, Suwon}
\affiliation{School of Physics, University of Sydney, NSW 2006}
\affiliation{Tata Institute of Fundamental Research, Mumbai}
\affiliation{Excellence Cluster Universe, Technische Universit\"at M\"unchen, Garching}
\affiliation{Toho University, Funabashi}
\affiliation{Tohoku Gakuin University, Tagajo}
\affiliation{Tohoku University, Sendai}
\affiliation{Department of Physics, University of Tokyo, Tokyo}
\affiliation{Tokyo Institute of Technology, Tokyo}
\affiliation{Tokyo Metropolitan University, Tokyo}
\affiliation{Tokyo University of Agriculture and Technology, Tokyo}
\affiliation{CNP, Virginia Polytechnic Institute and State University, Blacksburg, Virginia 24061}
\affiliation{Wayne State University, Detroit, Michigan 48202}
\affiliation{Yamagata University, Yamagata}
\affiliation{Yonsei University, Seoul}
  \author{X.~L.~Wang}\affiliation{Institute of High Energy Physics, Chinese Academy of Sciences, Beijing} 
         \affiliation{CNP, Virginia Polytechnic Institute and State University, Blacksburg, Virginia 24061} 
  \author{Y.~L.~Han}\affiliation{Institute of High Energy Physics, Chinese Academy of Sciences, Beijing} 
  \author{C.~Z.~Yuan}\affiliation{Institute of High Energy Physics, Chinese Academy of Sciences, Beijing} 
  \author{C.~P.~Shen}\affiliation{Graduate School of Science, Nagoya University, Nagoya} 
  \author{P.~Wang}\affiliation{Institute of High Energy Physics, Chinese Academy of Sciences, Beijing} 
  \author{I.~Adachi}\affiliation{High Energy Accelerator Research Organization (KEK), Tsukuba} 
  \author{H.~Aihara}\affiliation{Department of Physics, University of Tokyo, Tokyo} 
 \author{D.~M.~Asner}\affiliation{Pacific Northwest National Laboratory, Richland, Washington 99352} 
  \author{V.~Aulchenko}\affiliation{Budker Institute of Nuclear Physics SB RAS and Novosibirsk State University, Novosibirsk 630090} 
  \author{T.~Aushev}\affiliation{Institute for Theoretical and Experimental Physics, Moscow} 
% ...
\collaboration{The Belle Collaboration}

\date{\today}

\begin{abstract}

The cross section for $\EE\to \etajpsi$ between
$\sqrt{s}=3.8~\gev$ and $5.3~\gev$ is measured via initial state
radiation using 980~fb$^{-1}$ of data on and around the
$\Upsilon(nS)(n=1,2,3,4,5)$ resonances collected with the Belle
detector at KEKB. Two resonant structures at the $\psift$ and
$\psifto$ are observed in the $\etajpsi$ invariant mass
distribution. Fitting the mass spectrum with the coherent sum of
two Breit-Wigner functions, one obtains
$\BR(\psi(4040)\to\etajpsi)\cdot\Gamma_{\EE}^{\psi(4040)} =
(4.8\pm0.9\pm1.4)~\ev$ and
$\BR(\psi(4160)\to\etajpsi)\cdot\Gamma_{\EE}^{\psi(4160)} =
(4.0\pm0.8\pm1.4)~\ev$ for one solution and
$\BR(\psi(4040)\to\etajpsi)\cdot\Gamma_{\EE}^{\psi(4040)} =
(11.2\pm1.3\pm1.9)~\ev$ and
$\BR(\psi(4160)\to\etajpsi)\cdot\Gamma_{\EE}^{\psi(4160)} =
(13.8\pm1.3\pm2.0)~\ev$ for the other solution, where the first
errors are statistical and the second are systematic. This is the
first measurement of this hadronic transition mode of these two
states, and the partial widths to $\eta\jpsi$ are found to be
about $1~\mev$. There is no evidence for the $Y(4260)$, $Y(4360)$,
$\psi(4415)$, or $Y(4660)$ in the $\eta\jpsi$ final state, and
upper limits of their production rates in $\EE$ annihilation are
determined.

\end{abstract}


\maketitle

Many charmonium and charmoniumlike states have been discovered at
$B$-factories in the past decade. Some of these states are good
candidates for conventional charmonium states, while others
exhibit unusual properties consistent with expectations for exotic
states such as a multi-quark state, molecule, hybrid, or the
glueball~\cite{review}. In the vector sector, four exotic
charmoniumlike structures, $Y(4008)$ and $Y(4260)$ in $\EE\to
\ppjpsi$~\cite{belley,babay4260} and  $Y(4360)$ and $Y(4660)$ in
$\EE\to \pppsp$~\cite{pppsp,babay4324}, {have been} reported via
initial state radiation (ISR), in addition to the three known
excited $\psi$ states  above $4.0~\gevcs$: $\psift$, $\psifto$,
and $\psiftf$. It is unlikely that all seven of these states are
charmonia, as the potential models predict only five vector states
in this mass region~\cite{barnes}. The current understanding of
these states is based on limited statistics, and the fact that
some may be produced via mechanisms that are difficult to estimate
theoretically, such as final state rescattering~\cite{review},
makes the determination of which might be exotic even more
challenging. In order to further the understanding of the nature
of these states, it is important to investigate them using much
larger data samples.

An important study is the investigation of hadronic transitions
(either by an $\eta$ or a pion pair) between these states and a
lower charmonium state like the $\jpsi$. The CLEO collaboration
measured $\sigma(\EE\to\eta\jpsi)=15^{+5}_{-4}\pm 8$~pb at
$\sqrt{s} = 4120-4200~\mev$~\cite{cleo}, and the BESIII
collaboration reported $\sigma(\EE\to\eta\jpsi)=(32.1\pm 2.8)$~pb
at $\sqrt{s}=4009~\mev$~\cite{besiii}, which is in agreement with
the theoretical calculation including contributions from the known
$\psi$ states and the virtual charmed meson loops~\cite{zhaoq}.
However, the limited statistics of the CLEO analysis prevented the
measurement of the line shape of $\eta\jpsi$. Thus, it is
worthwhile to study the process $\EE\to \eta\jpsi$ via ISR with
the full Belle data sample to search for $\eta$ transitions from
these seven states to $\jpsi$. It is worth noting that the $\psi$
states are identified in decays to charmed meson pairs but not in
dipion transitions to lower $\psi$ states, while the opposite is
true of the $Y$ states. There may also be surprises from
transitions of unexpected states.

In this Letter, we report an investigation of the $\EE \to
\etajpsi$ process using ISR events observed with the Belle
detector~\cite{Belle} at the KEKB asymmetric-energy $e^+e^-$
collider~\cite{KEKB}. Here, $\jpsi$ is reconstructed in the
$\LL~(\ell=e,\mu)$ final state and $\eta$ in the $\GG$ and $\ppp$
final states. Due to the high background level from Bhabha
scattering, the $\jpsi\to \EE$ mode is not used in conjunction
with the decay mode $\eta\to \GG$. The integrated luminosity used
in this analysis is 980~fb$^{-1}$. About 70\% of the data were
collected at the $\Upsilon(4S)$ resonance, and the rest were taken
at other $\Upsilon(nS)$ ($n=1$, 2, 3, or 5) states or
center-of-mass (CM) energies a few tens of $\mev$ lower than the
$\Upsilon(4S)$ or the $\Upsilon(nS)$ peaks.

We use the {\sc phokhara} event generator~\cite{phokhara} to
simulate the process $\EE \to \gamma_{\rm ISR} \etajpsi$. In the
generator, one or two ISR photons may be emitted before forming
the resonance $X$, which then decays to $\etajpsi$, with $\jpsi\to
\EE$ or $\MM$ and $\eta\to\ppp$ or $\GG$.

For a candidate event, we require two (four) good charged tracks
with zero net charge for $\eta\to \GG$ ($\eta\to \ppp$). A good
charged track has impact parameters with respect to the
interaction point of $dr < 0.5$~cm in the $r$-$\phi$ plane and
$|dz|<5$~cm in the $r$-$z$ plane. The transverse momentum of the
leptons is required to be greater than $0.1~\gevc$. For each
charged track, information from different detector subsystems is
combined to form a likelihood for each particle species ($i$),
$\mathcal{L}_i$~\cite{pid}. Tracks with $\mathcal{R}_K =
\frac{\mathcal{L}_K}{\mathcal{L}_K+\mathcal{L}_\pi} < 0.4$ are
identified as pions with an efficiency of about 95\%, while 6\% of
kaons are misidentified as pions. Similar likelihood ratios are
formed for electron and muon identification~\cite{EID,MUID}. For
electrons from $\jpsi\to \EE$, both tracks are required to have
$\mathcal{R}_e > 0.1$. The bremsstrahlung photons detected in the
electromagnetic calorimeter (ECL) within 0.05 radians of the
original $e^+$ or $e^-$ direction are included in the calculation
of the $\EE(\gamma)$ invariant mass. For muons from $\jpsi\to
\MM$, one of the tracks is required to have $\mathcal{R}_\mu >
0.9$ and  the other track should have associated hits in the
$K_L$-and-muon detector (KLM) that agree with the extrapolated
trajectory of a charged track provided by the drift chamber. The
lepton ID efficiency is about 90\% for $\jpsi\to \EE$ and 87\% for
$\jpsi\to \MM$.

The $\eta$ is reconstructed from $\ppp$ and $\GG$ final states. To
reconstruct $\eta\to \ppp$, the $\pi^0$ is reconstructed from two
photons. A photon candidate is an ECL cluster with energy
$E(\gamma) > 25~\mev$ that does not match any charged tracks. The
mass resolution of $\pi^0$ is about $5.2~\mevcs$ from MC
simulation. Considering the low-mass tail, the invariant mass of
the photon pair is required to be between $110~\mevcs$ and
$150~\mevcs$ for a $\piz$ candidate. $\ppp$ combinations are
formed and are subject to a mass-constrained kinematic fit. When
there is more than one $\pi^0$ candidate, the combination with the
smallest $\chi^2$ from the mass-constrained fit is selected as the
$\eta$ candidate. Events with $\gamma$-conversions are removed by
requiring $\mathcal{R}_e < 0.75$ for the $\pip\pim$ tracks from
$\eta$ decays. In the reconstruction of $\eta\to\GG$ candidates,
two photon candidates are required with energies in the laboratory
frame satisfying $E(\gamma_l) > 0.15~\gev$ and $E(\gamma_h) >
0.4~\gev$, where the subscript $l$ ($h$) signifies the lower
(higher) energy photon.

The scatter plots of dilepton invariant mass $M_{\LL}$ versus
$\eta$-candidate invariant mass $M_{\ppp}$ or $\gamma_l\gamma_h$
invariant mass $M_{\GG}$ are shown in Fig.~\ref{mllmeta} for
events that survive these selection criteria. Here the invariant
masses are calculated with the momenta before the mass
constraints. A dilepton pair is considered as a $\jpsi$ candidate
if $M_{\LL}$ is within $\pm 45~\mevcs$ (the mass resolution being
$15~\mevcs$) of the $\jpsi$ nominal mass. The $\jpsi$ mass
sidebands are defined as $M_{\LL}\in [3.172,3.262]~\gevcs$ or
$M_{\LL}\in [2.932,3.022]~\gevcs$. A fit of the $M_{\ppp}$ or
$M_{\GG}$ distribution with a Gaussian plus a second-order
polynomial yields a mass resolution of $4.3~\mevcs$ for the
$\eta\to \ppp$ mode and $11.1~\mevcs$ for the $\eta\to \GG$ mode.
We define the $\eta$ signal region as $M_{\ppp} \in [0.5343,
0.5613]~\gevcs$ and $M_{\GG}\in [0.5,0.6]~\gevcs$ and the $\eta$
mass sideband regions as $M_{\ppp}\in [0.5748,0.6018]~\gevcs$ or
$M_{\ppp}\in [0.4938,0.5208]~\gevcs$, and $M_{\GG}\in
[0.35,0.45]~\gevcs$ or $M_{\GG}\in [0.65,0.75]~\gevcs$. The
central (surrounding) rectangles of Fig.~\ref{mllmeta} show the
$\eta\jpsi$ signal (sideband) regions. With $S1$ ($S2$)
representing the sum of the events in the four sideband boxes
nearest (diagonal) to the signal box, the normalization of the
sidebands is $S = 0.5\times S1 - 0.25\times S2$.

\begin{figure}[htbp]
 \psfig{file=mll-m3pi-draft.epsi,height=5.5cm}
 \psfig{file=mll-mgg-draft.epsi,height=5.5cm}
\caption{Invariant mass distributions of (a) $\LL$ vs. $\ppp$ and
(b) $\LL$ vs. $\GG$ for selected $\ppp\LL$ or $\GG\LL$ candidates
with invariant mass between $3.8~\gevcs$ and $5.3~\gevcs$. The box
in the center of each plot shows the $\etajpsi$ signal region
while the surrounding boxes show the sideband regions. }
\label{mllmeta}
\end{figure}

The detection of the ISR photon is not required; instead, we
require $-1~(\gevcs)^2 < \MMS < 2.0~(\gevcs)^2$, where $\MMS$ is
the square of the mass recoiling against the $\eta\jpsi$ system.
In calculating $\MMS$, the momenta of the $\jpsi$ and $\eta$ after
the kinematic fit are used to improve the resolution of $\MMS$.
The fit constrains signal candidates to the $\eta$ and $\jpsi$
masses, while events having $\eta$ or $\jpsi$ candidate masses
lying in sideband regions are fitted with masses constrained to
the center of the sideband region.

\begin{figure}[htbp]
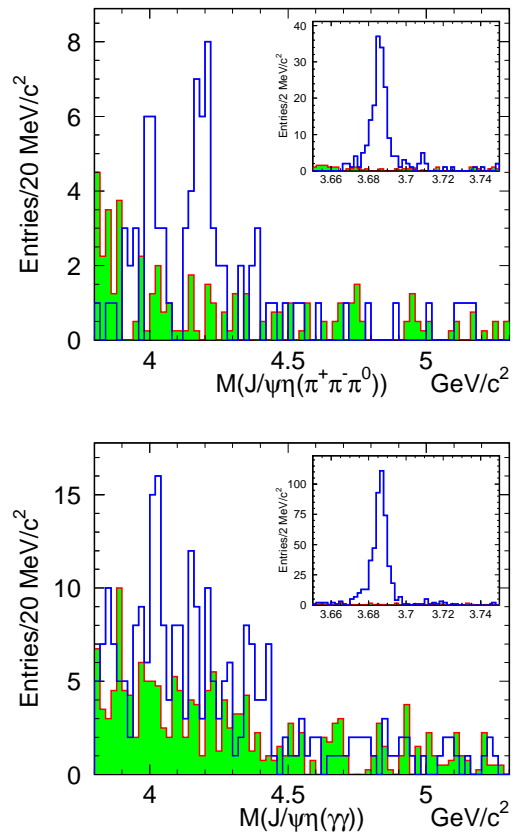

 \psfig{file=metajpsi-ppp-h-cf1.epsi,width=7.5cm}
 \psfig{file=mggjpsi-gg-h-cf1.epsi,width=7.5cm}
\caption{The invariant mass distribution of the $\eta\jpsi$
candidates. The top row shows the $\eta\to \ppp$ mode and the
bottom row shows the $\eta\to \GG$ mode. The open histograms are
from the $\eta$ and $\jpsi$ signal region, while the shaded ones
are from their sideband regions after the proper normalization.
The insets show the distributions around the $\psp$ mass region.}
\label{metajpsi}
\end{figure}

Figure~\ref{metajpsi} shows the $\etajpsi$ invariant mass
($M_{\eta\jpsi}$~\cite{metajpsi}) for selected candidate events,
together with background estimated from the scaled $\eta$ or
$\jpsi$ mass sidebands.  Two distinct peaks are evident in
Fig.~\ref{metajpsi}, one at $4.0~\gevcs$ and the other at
$4.2~\gevcs$, in addition to the dominant $\psp$ signal.
The cross section of $\EE\to \gamma_{\rm ISR}\psp$ in the full
Belle data sample is measured to be $13.9\pm 1.4~({\rm stat.})$~pb
using the $186\pm17$ $\eta\to \ppp$ events and $14.0\pm 0.8~({\rm
stat.})$~pb using the $470\pm25$ $\eta\to \GG$ events, in good
agreement with the production cross section of 14.7~pb calculated
by using the world average values of the mass, width, and partial
width to $\EE$ of $\psp$~\cite{PDG}, and the $\EE$ CM energies
correspond to the Belle data samples. Agreement between the
$\jpsi\to \EE$ and $\jpsi\to \MM$ modes is also observed.
The visible energy ($E_{\rm vis}$) and polar angle distributions
of the $\etajpsi$ system in the $\EE$ CM frame for the events with
$M_{\etajpsi}\in [3.8,5.3]~\gevcs$ agree well with the MC
simulation, confirming that the signal events are produced via
ISR.
Here, $E_{\rm vis}$ is the total energy of all final state photons
and charged particles. Charged particle energies are calculated
from track momenta assuming the tracks to be pions.

\begin{figure}[htbp]
 \psfig{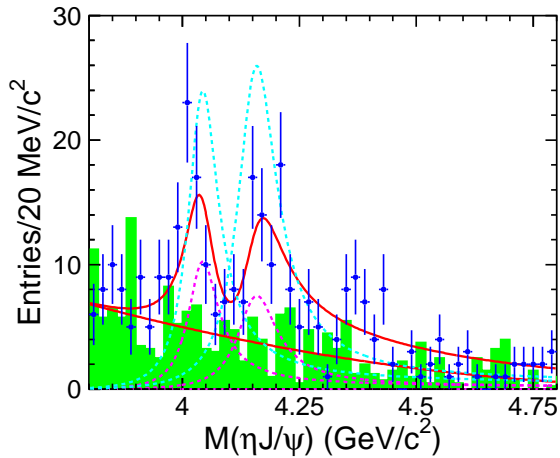}
\caption{The $\etajpsi$ invariant mass distribution and the fit
results. The points with error bars show the data while the shaded
histogram is the normalized $\eta$ and $\jpsi$ background from the
sidebands. The curves show the best fit on signal candidate events
and sideband events simultaneously and the contribution from each
Breit-Wigner component. The interference between the two
resonances is not shown. The dashed curves at each peak show the
two solutions (see text).} \label{fit}
\end{figure}

An unbinned maximum likelihood fit is performed to the mass
spectra $M_{\etajpsi}\in [3.8,4.8]~\gevcs$ from the  signal
candidate events and $\eta$ and $\jpsi$ sideband events
simultaneously, as shown in Fig.~\ref{fit}. The fit to the signal
events includes two coherent $P$-wave Breit-Wigner functions,
$BW_1$ for $\psift$ and $BW_2$ for $\psifto$, assuming that only
two resonances contribute to the $\eta\jpsi$ final states, and an
incoherent second-order polynomial background; the fit to the
sideband events includes  the same background function only. The
width of each resonance is assumed to be constant, and an overall
two-body phase-space factor is applied in the partial width to
$\etajpsi$. The signal amplitude is $M = BW_1+e^{i\phi}\cdot
BW_2$, where $\phi$ is the the relative phase between the two
resonances. In the fit, the BW functions are convolved with the
effective luminosity~\cite{kuraev} and $M_{\etajpsi}$-dependent
efficiency, which increases from 4\% at $M_{\etajpsi} =
4.0~\gevcs$ to 7\% at $M_{\etajpsi}=4.5~\gevcs$. The effect of
mass resolution, which is determined from MC simulation to be
$5-11~\mevcs$ over the resonant mass region, is small compared
with the widths of the observed structures, and therefore is
neglected. A fit performed with floating masses and widths for the
two structures yields a mass of $(4012\pm 5)~\mevcs$ and width of
$(54\pm 13)~\mev$ for the first, and a mass of $(4157\pm
10)~\mevcs$ and width of $(84\pm 20)~\mev$ for the second.
Their masses and widths are in agreement with those of the
$\psift$ and $\psifto$, and thus they are referred to hereafter as
the $\psift$ and $\psifto$. In the fit below, the masses and
widths of these two resonances are fixed to their world average
values~\cite{PDG} as the statistics are low here.

Figure~\ref{fit} and Table~\ref{two_sol}~\cite{beebf} show the fit
results. There are two solutions with equally good fit quality. To
determine the goodness of the fit, we bin the data (events in both
signal and sideband regions) so that the expected number of events
in a bin is at least seven and then calculate a $\chi^2/ndf$ of
71.4/46, corresponding to a confidence level (C.L.) of 0.9\%,
where $ndf$ is the number of degrees of freedom. The significance
of each resonance is estimated by comparing the likelihood of fits
with and without that resonance included. We obtain a statistical
significance of $6.5\sigma$ for $\psift$ and $7.6\sigma$ for
$\psifto$. Varying the masses and widths of resonances by
$1\sigma$, the fit range by $200~\mevcs$, and the order of the
background polynomial by one, we obtain a minimum statistical
significance of $6.0\sigma$ for $\psift$ and $6.5\sigma$ for
$\psifto$.

Taking $\Gamma_{\EE}^{\psi(4040)} = (0.86\pm 0.07)~\kev$ from
PDG~\cite{PDG}, one obtains $\BR(\psift\to \eta\jpsi)= (0.56\pm
0.10\pm 0.17)\%$ or $\BR(\psift\to \eta\jpsi)= (1.30\pm 0.15\pm
0.24)\%$; while using the PDG average value
$\Gamma_{\EE}^{\psifto} = (0.83\pm 0.07)~\kev$~\cite{PDG}, one
gets $\BR(\psifto\to \eta\jpsi) = (0.48\pm 0.10\pm 0.17)\%$ or
$(1.66\pm 0.16\pm 0.28)\%$. In each case, the first error is
statistical and the second is systematic. These indicate the
transition rates of these states to $\etajpsi$ are large, being of
order $1~\mev$.

Possible contributions from other excited charmonium(like) states
are examined. There is a cluster of events near the $M_{\eta\jpsi}
= 4.36~\gevcs$. Assuming it is the $Y(4360)$, the significance is
$1.1\sigma$ in a fit with the masses and widths of the $\psift$
and $\psifto$ fixed to their world average values~\cite{PDG}, or
$2.9\sigma$ if the masses and widths of $\psi(4040)$ and
$\psi(4160)$ are free. Besides the $Y(4360)$, the $Y(4260)$,
$\psi(4415)$ and $Y(4660)$ are in $[3.8, 5.3]~\gevcs$ mass region.
Fits that include each one of them and the masses and widths of
$\psift$ and $\psifto$ fixed to their world average
values~\cite{PDG} are performed to determine the upper limits of
$\BR\cdot\Gamma_{\EE}$. The systematic errors that will be
described later in the text together with those from the
uncertainties of the $\psi(4040)$ and $\psi(4160)$ resonant
parameters are considered in the upper limit determination. In
order to be conservative, the efficiencies have been lowered by a
factor of $1-\sigma_{sys}$ in the calculation. We obtain the upper
limits on $\BR(X\to\eta\jpsi)\cdot\Gamma^{X}_{\EE}$ for $X =
Y(4260)$, $Y(4360)$, $\psi(4415)$ and $Y(4660)$ are $14.2~\ev$,
$6.8~\ev$, $3.6~\ev$ and $0.94~\ev$ at 90\% C.L., respectively.

\begin{table}
\caption{Results of the fits to the $\etajpsi$ invariant mass
spectrum. The first errors are statistical and the second are
systematic. $M$, $\Gamma$, and $\BR\cdot \Gamma^{\psi}_{\EE}$ are
the mass (in $\mevcs$), total width (in $\mev$), product of the
branching fraction of $\psi\to\etajpsi$ and the $\psi\to\EE$
partial width (in $\ev$), respectively. $\phi$ is the relative
phase between the two resonances (in degrees).}\label{two_sol}
\begin{center}
\renewcommand{\arraystretch}{1.2}
\begin{tabular}{c c c }
\hline
     Parameters     & ~~Solution~I~~~ & ~~~Solution~II~~  \\\hline
 $M_{\psift}$        & \multicolumn{2}{c}{4039 (fixed)} \\
 $\Gamma_{\psift}$   & \multicolumn{2}{c}{80 (fixed)}  \\
 $\BR\cdot\Gamma_{\EE}^{\psift}$
                    & $4.8\pm0.9\pm1.4$ \quad & \quad $11.2\pm1.3\pm1.9$  \\
 $M_{\psifto}$       & \multicolumn{2}{c}{4153 (fixed)} \\
 $\Gamma_{\psifto}$  & \multicolumn{2}{c}{103 (fixed)}   \\
 $\BR\cdot\Gamma_{\EE}^{\psift}$
                    & $4.0\pm0.8\pm1.4$ \quad & \quad $13.8\pm1.3\pm2.0$  \\
 $\phi$             & $336\pm12\pm14$ & $251\pm4\pm7$ \\
 \hline
\end{tabular}\end{center}
\end{table}

To estimate the errors in $\BR\cdot \Gamma_{\EE}$, the
uncertainties from the choice of parametrization of the resonances
(especially introducing the mass dependence for the widths), the
masses and widths of resonances~\cite{PDG}, the fit range, the
background shape and the possible contributions from $\psp$ or
$\psi(4415)$ are considered. The total errors are 35.0\% and
14.8\% for solutions I and II, respectively. The particle ID
uncertainty is 5.5\%; the uncertainty in the tracking efficiency
is 0.35\% per track and is additive; the uncertainty in the photon
reconstruction is 2\% per photon. The uncertainties in the $\jpsi$
mass, $\eta$ mass, and $\MMS$ requirements are measured with the
control sample $\EE\to \psp\to \etajpsi$. The efficiencies of the
requirements on the data are obtained from the fits of the
corresponding distributions. The MC efficiency is found to be
higher than in data by $(2.3\pm 2.6)\%$ for the $\ppp$ mode and
$(0.1\pm 1.6)\%$ for the $\GG$ mode. A correction factor 1.023 is
applied to the $\ppp$ final state, and 2.6\% is conservatively
taken as the associated systematic error of the sum for $\ppp$ and
$\GG$ modes.

Belle measures luminosity with 1.4\% precision while the
uncertainty of the generator {\sc phokhara} is less than
1\%~\cite{phokhara}. The trigger efficiency for the events
surviving the selection criteria is around 91\% with an
uncertainty smaller than 2\%.
The uncertainties in the intermediate decay branching fractions
taken from Ref.~\cite{PDG} contribute a systematic error of less
than 1.6\%. The statistical error in the MC determination of the
efficiency is 0.2\%.

Assuming all the sources are independent and adding them in
quadrature, we obtain total systematic errors in $\BR\cdot
\Gamma_{\EE}$ of 36\% for Solution~I and 17\% for Solution~II for
both $\psift$ and $\psifto$.

\begin{figure}[htbp]
 \psfig{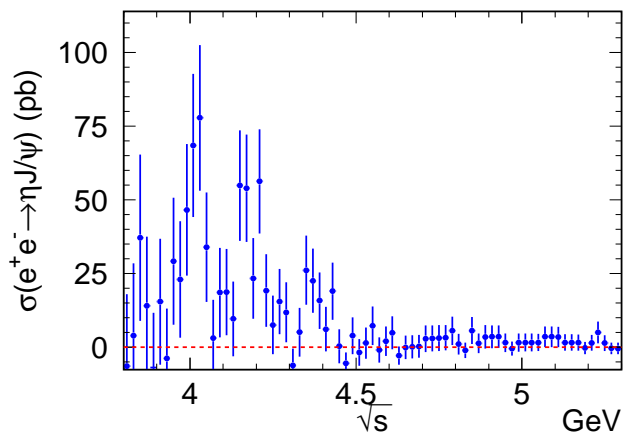}
\caption{The measured $\EE\to \etajpsi$ cross section for
$\sqrt{s}=3.8~\gev$ to $5.3~\gev$. The errors are the summed
statistical errors of the numbers of signal and background events.
A systematic error of 8.0\% common to all the data points are not
shown.} \label{xs_full}
\end{figure}

The cross section for $\EE\to \etajpsi$ for each $\etajpsi$ mass
bin is calculated according to
\[
 \sigma_i = \frac{n^{\rm obs}_i - n^{\rm bkg}_i}
  { \lum_i\times\sum\limits_{j}\eff_{ij}\BR_j},
\]
where $j$ is the $j$-th mode of $\eta\jpsi$ decays ($j=\ppp\EE$,
$\ppp\MM$, and $\GG\MM$); $n^{\rm obs}_i$, $n^{\rm bkg}_i$,
$\eff_{ij}$, $\lum_i$, and $\BR_j$ are number of events observed
in data, number of background events estimated from sidebands,
detection efficiency of the $j$-th mode, effective luminosity in
the $i$-th $\eta\jpsi$ mass bin, and the branching fraction of
$\eta\jpsi$ decays into the $j$-th mode~\cite{PDG}, respectively.
The resulting cross sections in the full solid angle are shown in
Fig.~\ref{xs_full}, where the error bars include the statistical
uncertainties in the signal and the background subtraction. The
systematic error for the cross section measurement, which
includes all the sources that have been described 
other than those arising from the details of the fit to the mass
spectrum, is 8.0\% and common to all the data points. The cross
sections of $\EE\to \etajpsi$ are around 70~pb and 50~pb at the
$\psift$ and $\psifto$ peaks, respectively, to be compared with
around 20~pb and 10~pb measured in $\EE\to \ppjpsi$~\cite{belley}.

In summary, the $\EE\to\etajpsi$ cross section is measured from
$3.8~\gev$ up to $5.3~\gev$ for the first time. Two distinct
resonant structures, the $\psift$ and $\psifto$, are observed.
This is the first time that the $\psift$ and $\psifto$ have been
observed to decay to final states not involving charm meson pairs.
The products of the branching fraction to $\etajpsi$ and the $\EE$
partial width are determined to be
$\BR(\psi(4040)\to\etajpsi)\cdot\Gamma_{\EE}^{\psift} =
(4.8\pm0.9\pm1.4)~\ev$ and
$\BR(\psi(4160)\to\etajpsi)\cdot\Gamma_{\EE}^{\psifto} =
(4.0\pm0.8\pm1.4)~\ev$ for one solution; or
$\BR(\psi(4040)\to\etajpsi)\cdot\Gamma_{\EE}^{\psift} =
(11.2\pm1.3\pm1.9)~\ev$ and
$\BR(\psi(4160)\to\etajpsi)\cdot\Gamma_{\EE}^{\psifto} =
(13.8\pm1.3\pm2.0)~\ev$ for the other solution. These transition
rates correspond to about $1~\mev$ partial widths to $\etajpsi$
for these two states. We find no evidence for the $Y(4260)$,
$Y(4360)$, $\psi(4415)$ or $Y(4660)$ in the  $\eta\jpsi$ final
states, and upper limits of their production rates in $\EE$
annihilation are determined. The present measurement reveals clear
peaks due to the $\psift$ and $\psifto$ decays observed in
experimental data that are absent in the prediction in
Ref.~\cite{zhaoq}, although the theoretical calculation with
carefully chosen parameters agrees with the measured cross
sections of $\EE\to\eta\jpsi$.

We thank the KEKB group for excellent operation of the
accelerator; the KEK cryogenics group for efficient solenoid
operations; and the KEK computer group, the NII, and PNNL/EMSL for
valuable computing and SINET4 network support. We acknowledge
support from MEXT, JSPS and Nagoya's TLPRC (Japan); ARC and DIISR
(Australia); NSFC (China); MSMT (Czechia); DST (India); INFN
(Italy); MEST, NRF, GSDC of KISTI, and WCU (Korea); MNiSW
(Poland); MES and RFAAE (Russia); ARRS (Slovenia); SNSF
(Switzerland); NSC and MOE (Taiwan); and DOE and NSF (USA).


\begin{thebibliography}{**}

\bibitem{review} For a recent review, see N.~Brambilla {\it et al.},
\Journal\EPJC{71}{1534}{2011}.

\bibitem{belley} C.~Z.~Yuan {\it et al.} (Belle Collaboration),
\Journal\PRL{99}{182004}{2007}.

\bibitem{babay4260} B.~Aubert {\it et al.} (BaBar Collaboration),
\Journal\PRL{95}{142001}{2005}; J. P. Lees {\it et al.} (BaBar Collaboration),
\Journal\PRD{86}{051102}{2012}.

\bibitem{pppsp} X.~L. Wang {\it et al.} (Belle Collaboration),
\Journal\PRL{99}{142002}{2007}.

\bibitem{babay4324} B.~Aubert {\it et al.} (BaBar Collaboration),
\Journal\PRL{98}{212001}{2007}.

\bibitem{barnes} S.~Godfrey and N. Isgur,
\Journal\PRD{32}{189}{1985}; T.~Barnes, S. Godfrey and E. S. Swanson,
\Journal\PRD{72}{054026}{2005}; G.~J.~Ding, J. J. Zhu and M. L. Yan,
\Journal\PRD{77}{014033}{2008}.

\bibitem{cleo} T.~E.~Coan {\it et al.} (CLEO Collaboration),
\Journal\PRL{96}{162003}{2006}.

\bibitem{besiii} M.~Ablikim {\it et al.} (BESIII Collaboration),
\Journal\PRD{86}{071101}{2012}.

\bibitem{zhaoq} Q.~Wang, X.~H.~Liu and Q.~Zhao,
\Journal\PRD{84}{014007}{2011}.

\bibitem{Belle} A.~Abashian {\it et al.} (Belle Collaboration),
\Journal\NIMA{479}{117}{2002}.

\bibitem{KEKB} S.~Kurokawa and E.~Kikutani,
\Journal\NIMA{499}{1}{2003}
and other papers included in this volume.

\bibitem{phokhara} G.~Rodrigo {\it et al.},
\Journal\EPJC{24}{71}{2002}. For a review on the generator, see:
S.~Actis {\it et al.}, \Journal\EPJC{66}{585}{2010}.


\bibitem{pid} E.~Nakano,
\Journal\NIMA{494}{402}{2002}.


\bibitem{EID} K.~Hanagaki {\it et al.},
\Journal\NIMA{485}{490}{2002}.

\bibitem{MUID} A.~Abashian {\it et al.},
\Journal\NIMA{491}{69}{2002}.

\bibitem{metajpsi}
$M_{\etajpsi} = M_{\ppp\LL} - M_{\ppp} - M_{\LL} + m_{\eta} + m_{\jpsi}$
for the $\eta\to \ppp$ mode and $M_{\etajpsi} = M_{\GG\LL} -
M_{\GG} - M_{\LL} + m_{\eta} + m_{\jpsi}$ for the $\eta\to \GG$
mode, where $m_{\eta}$ and $m_{\jpsi}$ are the nominal $\eta$ and
$\jpsi$ masses, respectively.


\bibitem{PDG} J.~Beringer {\it et al.} (Particle Data Group),
\Journal\PRD{86}{010001}{2012}.

\bibitem{kuraev} E.~A.~Kuraev and V.~S.~Fadin,
Sov. J. Nucl. Phys.  {\bf 41}, 466 (1985) [Yad. Fiz. {\bf 41}, 733
(1985)].

\bibitem{beebf}
Fitting the $M_{\eta\jpsi}$ spectrum with the product
$\BR(\psi\to\etajpsi)\cdot \BR(\psi\to\EE)$ as a parameter, and
the masses and widths of $\psift$ and $\psifto$ fixed to world
average values~\cite{PDG}, we obtain \(\BR(\psi\to\etajpsi)\cdot
\BR(\psi\to\EE) = (5.1\pm 1.4\pm 1.4)\times 10^{-8}\) and
\((2.8\pm 0.9\pm 0.9)\times 10^{-8}\) for the $\psift$ and
$\psifto$, respectively, for solution I; and
\(\BR(\psi\to\etajpsi)\cdot \BR(\psi\to\EE) = (12.8\pm 2.1\pm
1.7)\times 10^{-8}\) and \((12.8\pm 1.7\pm 1.9)\times 10^{-8}\)
for the $\psift$ and $\psifto$, respectively, for solution II.

\end{thebibliography}
\end{document}